\documentclass[a4paper,twocolumn,11pt]{quantumarticle}
\pdfoutput=1
\usepackage[utf8]{inputenc}
\usepackage[english]{babel}
\usepackage[T1]{fontenc}
\usepackage{amsmath}
\usepackage{hyperref}
\usepackage{subcaption}
\usepackage{tikz}
\usepackage{lipsum}

\begin{document}

\title{Parallel Quantum Hough Transform}
\author{Frank Klefenz}
\affiliation{Fraunhofer Institute for Digital Media Technology IDMT, 98693 Ilmenau, Germany}
\orcid{0000-0002-7994-3444}
\email{frank.klefenz@idmt.fraunhofer.de}
\homepage{https://www.idmt.fraunhofer.de}
\author{Nico Wittrock}
\affiliation{International Iberian Nanotechnology Laboratory INL, 4715-330 Braga, Portugal}
\author{Frank Feldhoff}
\affiliation{Advanced Electromagnetics Group, Ilmenau University of Technology, 98693 Ilmenau, Germany}

%\thanks{You can use the \texttt{\textbackslash{}email}, \texttt{\textbackslash{}homepage}, and \texttt{\textbackslash{}thanks} commands to add additional %information for the preceding \texttt{\textbackslash{}author}. If applicable, this can also be used to indicate that a work has previously been published %in conference proceedings.}

\texttt{unpublished}
%\texttt{noarxiv}

\maketitle

%\begin{abstract}
%  In the standard, \texttt{twocolumn}, layout the abstract is typeset as a bold face first paragraph.
%  Quantum also supports a \texttt{onecolumn} layout with the abstract above the text.
%  Both can be combined with the \texttt{titlepage} option to obtain a format with dedicated title and abstract pages that are not included in the page %count.
%  This format can be more suitable for long articles.
% The \texttt{abstract} environment can appear both before and after the \texttt{\textbackslash{}maketitle} command and calling %\texttt{\textbackslash{}maketitle} is optional, as long as there is an \texttt{abstract}.
%  Both \texttt{abstract} and \texttt{\textbackslash{}maketitle} however must be placed after all other \texttt{\textbackslash{}author}, %\texttt{\textbackslash{}affiliation}, etc.\ commands, see also Section~\ref{sec:title-information}.
%  If you provide the ORCID number of an author by using the \texttt{\textbackslash{}orcid} command, the author name becomes a link to their page on %\href{http://orcid.org/}{orcid.org}.
%\end{abstract}

\begin{abstract}
Few of the known quantum algorithms can be reliably executed on a quantum computer. 
Therefore, as an extension, we propose a Parallel Quantum Hough transform (PQHT) algorithm that we execute on a quantum computer. We give its implementation and discuss the results obtained. The PQHT algorithm is conceptually divided into a parallel rotation stage consisting of a set of connected programmable \texttt{RZ} rotation gates, with adjustable node connections of coincidence detectors realized with quantum logic gates. The modules were developed using IBM Quantum Composer and tested using the IBM QASM simulator. Finally, the modules were programmed using the Python package Qiskit and the jobs were sent to distributed IBM Q System One quantum computers. The successful run results on Fraunhofer Q System One in Ehningen will be presented as a proof of concept for the PQHT algorithm.
\end{abstract}

%In the \texttt{twocolumn} layout and without the \texttt{titlepage} option a paragraph without a previous section title may directly follow the abstract.
%In \texttt{onecolumn} format or with a dedicated \texttt{titlepage}, this should be avoided.

\section{Introduction}
%\label{sec:introduction}

The Hough transform is an image processing technique for finding mathematically predefined objects such as lines and circles in an image \cite{ballard1981generalizing}.
A survey of applications and implementations are given in \cite{mukhopadhyay2015survey}.
It is actually used in autonomous driving as for instance for traffic light and parking lot recognition and lane detection 
\cite{gautam2023image, rahman2020analysis, zakaria2023lane}. The Hough transform is also applied in event-based neuromorphic vision \cite{seifozzakerini2018hough}. Its state of the art applications are robot drilling, 3D reconstruction of buildings and control of UAV flights \cite{ayyad2023neuromorphic, bachiller2018spiking, vitale2021event}. The neuromorphic approach is supported by neuro-physiological studies of a cell model in the ventral visual pathway for the detection of circles of curvature constituting figures \cite{kawakami2020cell}. 
All these applications are to be executed in real time. Hence the Hough transform needs execution speed-ups. This is achievable by massive parallelisation of the Hough transform algorithm. The architectural design options for parallelisation are discussed in \cite{laghari2012processor}. Its execution speed has been accelerated in various GPU and FPGA implementations \cite{park2019lane, bailey2017streamed}.
We adopt the parallel Hough algorithm expressed in a systolic array form, which has been developed by R. Lay and F. Klefenz for detecting particle tracks in the OPAL drift chamber of the Large Electron-Positron (LEP) collider ring at CERN, Switzerland, for its quantum adaptation 
\cite{256642, DBLP:phd/dnb/Klefenz92}. We translate it step by step into its quantum form. We took that decision because of its high grade of parallelisation, which led to its high execution speed. The systolic array version is able to operate many parallel input data streams. It continuously outputs the Hough feature space without any queuing or scheduling conflicts. A second argument is that this parallel Hough transform algorithm has been implemented and yet run in FPGA boards and Asics \cite{Klefenz91asystolic}. The Asic version was realized by A. Epstein in 2002 \cite{1003733}. Its performance has been tested through the high detection rate of found particle tracks in the OPAL drift chamber of the Large Electron-Positron (LEP) collider ring at CERN in 1992 \cite{256642}. Its operation time is less than 4 $\mu$s and fulfills the real-time criteria. 
The hardware concept of the parallel Hough transform is a systolic array consisting of \textit{n} individually programmable shift/delay lines and \textit{m} coincidence counters that recognize up to \textit{m} patterns in parallel \cite{Klefenz91asystolic}. The shift/delay lines are connected by an input register at their initiating sites activating the signal flows, if triggering pixels from the sense wires are set. A pre-calculated cascade of orchestrated shift/delay sequences in the systolic array performs the Hough transform. The Hough feature space is limited by a range between minimum/maximum curvature for circles or minimum/maximum slope for straight lines. The mathematical Hough transform equations are numerically mapped to this range and fitted to the fixed grid with $n$ shift/delay lines and $m$ features in Hough space. A Hough array fitting program pre-calculates the shift/delay sequences. The FPGA boards and the Asic operate on 35 binary input pixel streams and outputs 32 different Hough features each clock cycle.

\section{Parallel Quantum Hough Transform (PQHT)}
Quantum image processing is an emerging research field and its potential applications are openly discussed \cite{wang2021review}. Images are represented by ordered sets of pixels with assignment of pixel intensities \cite{yan2022toward, amankwah2022quantum}. Varadajan and Ma et al. give proposals for a quantum Hough transform and quantum Radon transforms \cite{varadarajan2013quantum,ma2021quantum}.
%The polar coordinate system $(r,\phi)$ is transformed to a Cartesian $(x,y)$ coordinate system with square pixels of equal size for better visualization. This is done by drawing circular arcs with starting angle $\phi_{s}$ and radius of curvature $r_{c}$ or straight lines with different slopes $a$ initiated from the common origin (0,0) in polar coordinates and (0,0) in the Cartesian coordinate system. By applying Bresenham's algorithm every Cartesian pixel, which is swept by an arc or line, is set to 1\cite{pitteway1980bresenham}.
Our proposed parallel quantum Hough transform executes the same way as the systolic parallel Hough transform by using the same topology and processing steps. 
Each flip-flop is substituted by a controlled \texttt{RZ} rotation in the quantum case, expressing the shift/delay lines.
%In the quantum case, the shift/delay lines are expressed by concatenations of controlled \texttt{RZ} rotations. 
The coincidence detection units are expressed by quantum logic circuits. 
%The systolic parallel Hough transform is executed in an array of flip-flop elements. Each flip-flop is substituted by a controlled \texttt{RZ} rotation.
We take the IBM Q System One as a test system. Since the IBM Q System One has only 27 qubits available, we have to reduce as a restriction the 35 input streams $\times$ 32 Hough features to a manageable size of $3\times4$ for the PQHT.
In the systolic FPGA version, images are fed column by column into the systolic Hough processing array, whereas in  a quantum computer all image pixels must be stored at once at initialization as there is no FIFO mechanism to load an image column by column via bidirectional memory registers. In the quantum version, multiple consecutive image columns are aligned to an input string vector. To preserve the image pixel array topology, a special pixel indexing scheme in the rotational space has been chosen. All binary pixels in the input vector are set according to their connectivity relation at ordered positions in the rotational space by increments of \texttt{RZ}.
%The pixel position in x direction is given by \texttt{RZ}. The \texttt{RZ} rotational space is discretized by allowing only hop sizes of $n\cdot4\pi$. The \textit{y} position of the pixel is given by the number of the Qubit-line, and the \textit{x} position by the number $n$ of rotation steps of $n\cdot4\pi$. 
%A binary pixel $i(0,0)$ is assigned to Qubit line 0 for the \textit{y} coordinate value $0$ and set to the starting rotational position $-\pi$ for the \textit{x} coordinate value $0$. The binary pixels $i(0,1)$ and $i(0,2)$ are assigned to Qubit line 1 and Qubit line 2 for the \textit{y} coordinate value $0$ and are set to the starting rotational positions $-\pi$ for the \textit{x} coordinate value $0$. The $x$ values of the second column are positioned at $-5\pi$ and the pixels of the third column at $-9\pi$ and the $y$ value determines the Qubit line. 
The input pixels of our consideration $3 \times 3$ pixel grid are labelled acoording to \textbf{Fig.}~\ref{fig:labelling_qbits}. 
\begin{figure}[ht!]
    \centering
    \begin{tabular}{|c|c|c|}
    \hline
     %$(2, -9\pi)$ &$(2, -5\pi)$  &$(2, -\pi)$  \\ \hline 
     %$(1, -9\pi)$ & $(1, -5\pi)$ & $(1, -\pi)$ \\
     %\hline
     %$(0, -9\pi)$ & $(0, -5\pi)$ & $(0, -\pi)$ 
      $(-9\pi, 2)$ &$(-5\pi, 2)$  &$(-\pi, 2)$  \\ \hline 
      $(-9\pi, 1)$ & $(-5\pi, 1)$ & $(-\pi, 1)$ \\
    \hline
      $(-9\pi, 0)$ & $(-5\pi, 0)$ & $(-\pi, 0)$      
      \\ \hline
\end{tabular}
    \caption{Labelling of the input pixels}
\label{fig:labelling_qbits}
\end{figure}
%Each vector element is instantiated with its pixel value as for instance $i(0,0)$ either 0 or 1 for the binary case by setting the input to $| 0\rangle$, if the pixel value is $0$, or setting the pixel position to $| 1\rangle$, if the pixel value is $1$. This is done by placing the pixels with value $1$ at odd starting positions $n\cdot4\pi-\pi$ and pixels with value $0$ to even positions $n\cdot4\pi$.
%The smallest Hough feature space is a $2\times2$ input pixel array. Only two patterns are distinguishable, either a $90^\circ$ vertical bar or a $45^\circ$ diagonal bar.
We decided to demonstrate the PQHT for an image of size $3\times3$ pixels for straight lines with different slopes in an angular intercept from $45^\circ$ to $90^\circ$ for simplicity, as seen in \textbf{Fig.}~\ref{fig:fig1-input-pattern}. % In the $3\times3$ grid, 
There are four distinguishable patterns: $90^\circ$ vertical bar,  $75^\circ$ bar, $60^\circ$ bar, $45^\circ$ diagonal bar.
The pixels are indexed with the first column at the right, the second in the middle and the third at the left. The starting pixel in the origin is at the lower right at $(-\pi,0)$.
A vertical bar is represented by the set of pixels $[(-\pi,0)$, $(-\pi,1)$, $(-\pi,2)]$, a $75^\circ$ bar by $[(-\pi,0)$, $(-\pi, 1)$, $(-5\pi,2)]$, a $60^\circ$ bar by $[(-\pi,0)$, $(-5\pi,1)$, $(-5\pi,2)]$, the diagonal bar $45^\circ$ by $[(-\pi,0)$, $(-5\pi,1)$, $(-9\pi,2)]$ (see \textbf{Fig.}~\ref{fig:fig1-input-pattern}).

\begin{figure}[!ht]
    \centering
     \includegraphics[width=\columnwidth]{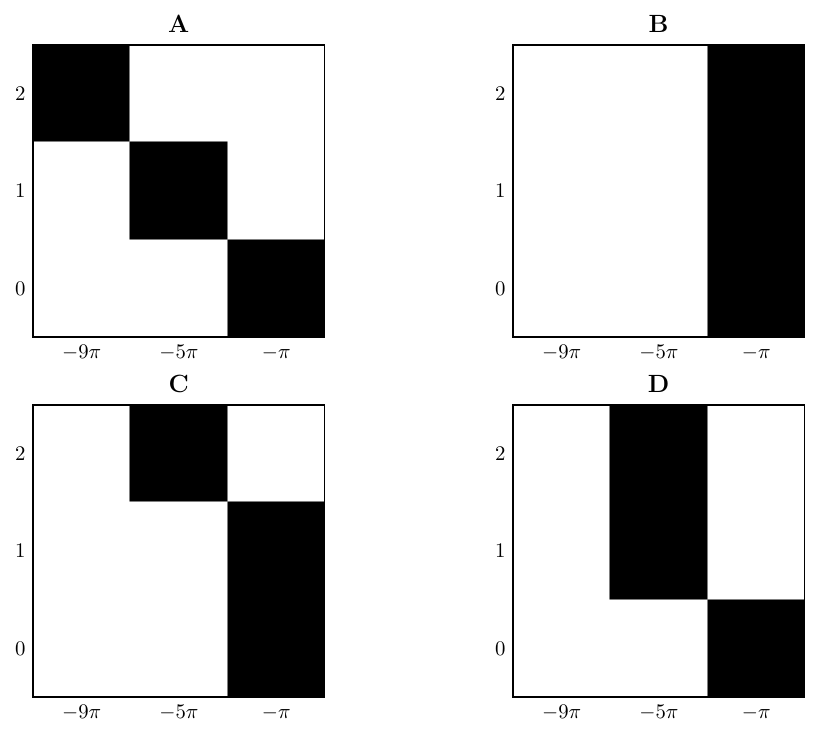}
     \caption{Different monochromatic input patterns with different angles are used to test the circuit. Black pixels represent a '1' and white pixels a '0'; \textbf{A} : 45$^{\circ}$, \textbf{B} : 90$^{\circ}$, \textbf{C} : 75$^{\circ}$ and \textbf{D} : 60$^{\circ}$.}  
     \label{fig:fig1-input-pattern}
\end{figure}

The quantum systolic array is built with Qubit lines consisting of concatenated \texttt{RZ} operators and columnwise intertwined coincidence detectors.
An initial Hadamard gate initiates, a controlled \texttt{RZ} rotation gate performs, and a closing Hadamard gate ends a rotation cycle.  \texttt{RZ} rotations performed in steps sizes of $n\cdot4\pi$ result in unique $| 1\rangle$ and $| 0\rangle$ states depending on the initial start positions at odd or even $-\pi$ positions. An odd position $n\cdot4\pi-\pi$ is in state $| 1\rangle$, an even position $n\cdot4\pi$ is in state $| 0\rangle$. Successive \texttt{RZ} rotations from left to right perform the shift/delay operations, which are tuned separately for each Qubit line. Boolean quantum logic applies as \texttt{RZ} rotations by $n\cdot4\cdot\pi$ always result in a Qubit state of either $| 1\rangle$ or $| 0\rangle$. A coincidence unit is dedicated to each Hough feature. A maximum-finder circuit realizes the coincidence unit. The coincidence is realized by ANDing the output of Qubit 0 and Qubit 1 and ANDing the result with Qubit line 2. If the three inputs are set to 1, a maximum event has been registered. The AND is realized with the Toffoli (CCNOT) gate.

\begin{figure*}[ht!]
    \centering
    \includegraphics[width=\linewidth]{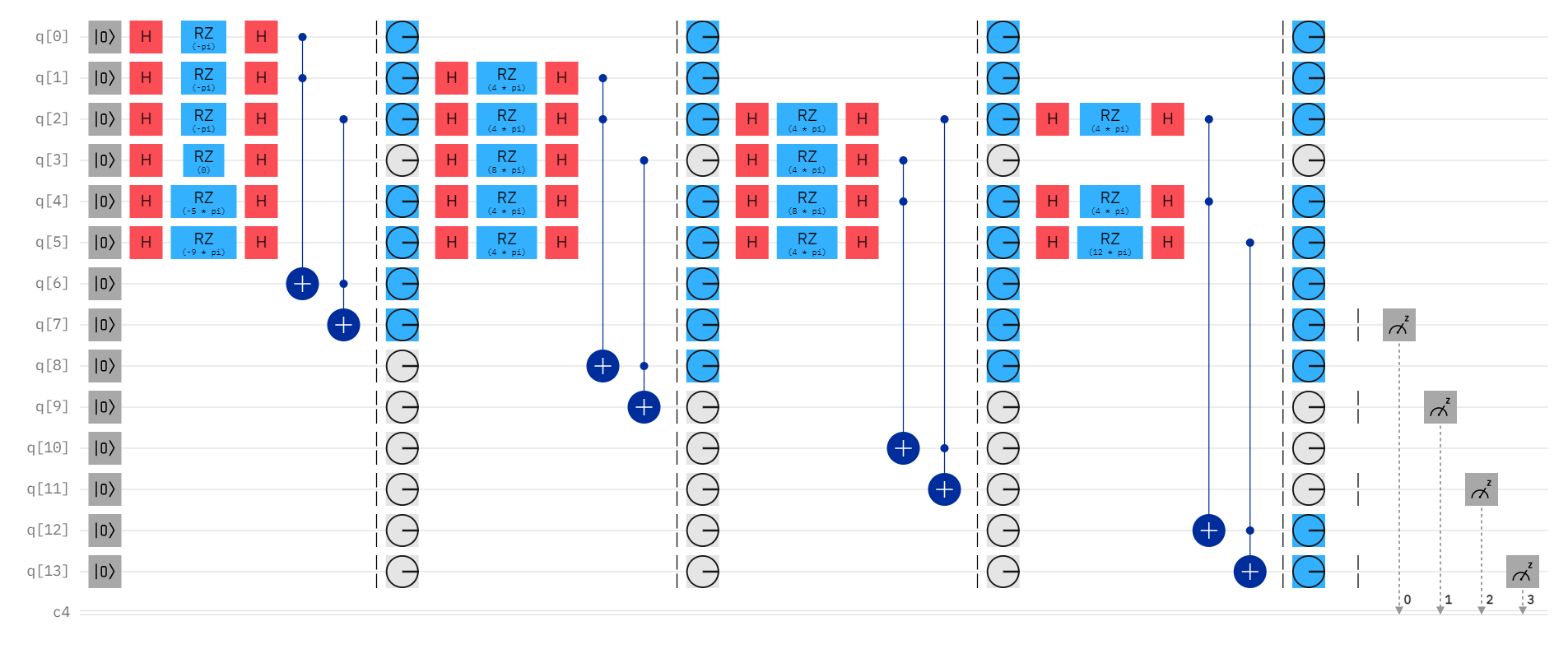}
    \caption{A separate Qubit line is assigned to each pixel. The vertical bar and the diagonal bar are set by initially placing the pixels to their initial odd starting positions $-n\cdot4\pi-\pi$ and the pixel with value 0 to the common default $\pi=0$ in the first block at the left. The vertical bar (90$^{\circ}$) is detected in the first coincidence unit with detector output q(7) blue and the diagonal bar in the fourth column with q(13) blue.}
    \label{fig:PQHTfirstcolumn}
\end{figure*}

\section{The Quantum Cycle}
The PQHT for straight lines in quantum rotation space \texttt{RZ} with discrete hop sizes $n\cdot4\pi$ for a $3\times3$ pixel grid is given. 
As shown in \textbf{Fig.}~\ref{fig:fig1-input-pattern}, we only consider four input patterns, namely those straight lines that 
cross the bottom-right edge and
are above the diagonal. 
All other lines can be obtained by mirroring.
%Only pixels in the image field above and along the diagonal are considered, and the three pixels below the diagonal in the left lower corner remain unplugged. 
In the Quantum circuit of \textbf{Fig.}~\ref{fig:PQHTfirstcolumn},  
q(0-5) are the rotation lines; to the lowest pixel in first image column at the right $(-\pi, 0)$ Qubit(0) is assigned,  Qubit(1) to $(-\pi,1)$, Qubit(2) $(-\pi,2)$, Qubit(3) $(-5\pi,2)$ Qubit(4)$(-5\pi,1)$, Qubit(5) $(-9\pi,2)$. The Qubits q(0-5) are placed to their rotation positions by \texttt{RZ} rotations, $-n\cdot4\pi-\pi$ whereas $n$ is the row number counted from right to left. The first image column on the right side is set to rotation position $-\pi$, the second column to $-5\pi$, the third to $-9\pi$. As the phases are all odd, all Qubits are in state $| 1\rangle$. The program starts with  \texttt{RZ} parallel rotations enclosed in common Hadamard blocks.  After each block closing Hadamard a maxfinder quantum boolean circuit composed of two Toffoli gates is inserted. Exemplarily, for the first maxfinder block, the first Toffoli logically ANDs q(0) and q(1) and outputs the result in carry line q(6). The second Toffoli ANDs q(2) with the carry Qubit line q(6) and outputs the final result in q(7). Therefore, q(7) signals if a maximum condition has been found.

\section{Step-by-step passage of the circuit}
The parallel quantum Hough transform algorithm has been developed graphically with the IBM Quantum Composer. A proof of concept is given for a vertical bar and a diagonal bar (see \textbf{Fig.} \ref{fig:PQHTfirstcolumn}). The inserted clock-like inline phase disks allow observing the Qubit states and phases at each pattern recognition unit. The final result states and phases are shown in the last inline phase disk slice on the right. With the inspector, the actual Qubit states and their phases are observable step by step. 
We explain our quantum circuit by taking the example of \textbf{Fig.}~\ref{fig:PQHTfirstcolumn}. 
The Qubits q(0-2), q(4-5) start with $| 1\rangle$ as they are placed at odd $-\pi$ positions. Pixel q(3) is white and hence initialized by \texttt{RZ(0)}, so that it rests inert and is always in state $| 0\rangle$ in the phase disk slices. The opening Hadamards put q(0-5) into superposition states. The following \texttt{RZ} in q(0-2), q(4-5) flip the states of q(0-2) and q(4-5) always to consecutive $n\cdot4\cdot\pi-\pi$ end positions so that the states are always $| 1\rangle$. The end-stop Hadamards kickback q(0-2), q(4-5) to $| 1\rangle$ and q(3) always remains in state $| 0\rangle$. 
%After the end Hadamards 
Finally, boolean quantum logical operations are applied. The first Toffoli in the first Boolean block sets the carry line q(6) to state $| 1\rangle$, if the AND condition is true. The second Toffoli flips q(7) to state $| 1\rangle$ if the AND condition carry q(6) and q(3) is true. Qubit q(7) in state $| 1\rangle$ signals that a vertical bar has been found. Qubit q(13) in state $| 1\rangle$ in the last phase disk slice indicates, that the diagonal pattern has been found. 

\section{Circuit verification}
\begin{figure*}[!ht]
    \centering
    \begin{subfigure}[b]{0.49\linewidth}
         \centering
         \includegraphics[width=\textwidth]{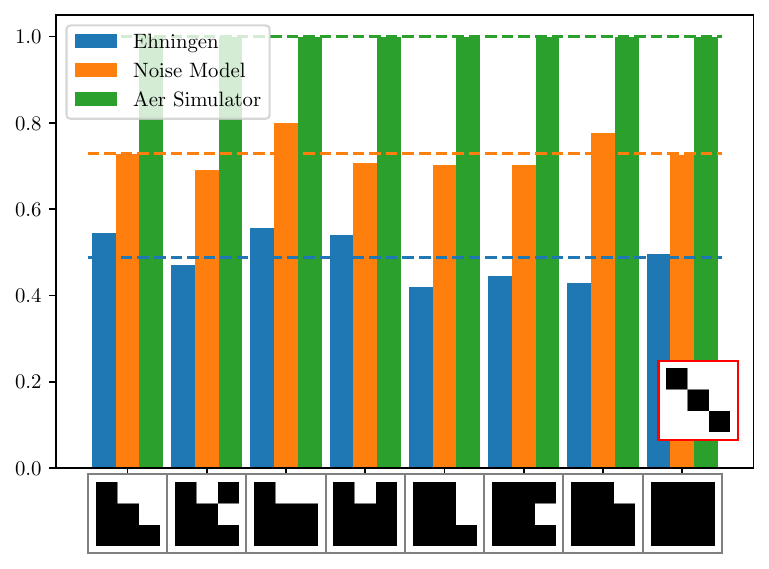}
         \caption{}
         \label{fig:result_plot_from_all_simulations:A}
     \end{subfigure}
     \begin{subfigure}{0.49\linewidth}
         \centering
         \includegraphics[width=\textwidth]{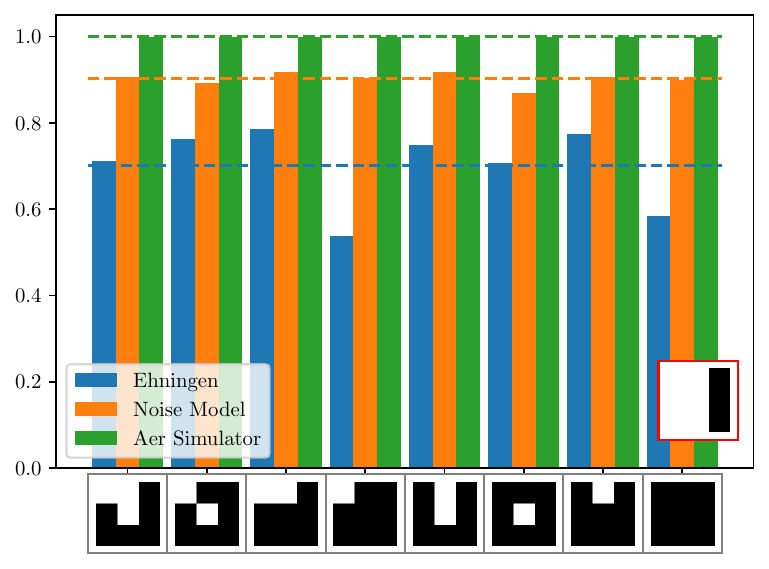}
         \caption{}
         \label{fig:result_plot_from_all_simulations:B}
     \end{subfigure}
    \begin{subfigure}[b]{0.49\linewidth}
         \centering
         \includegraphics[width=\textwidth]{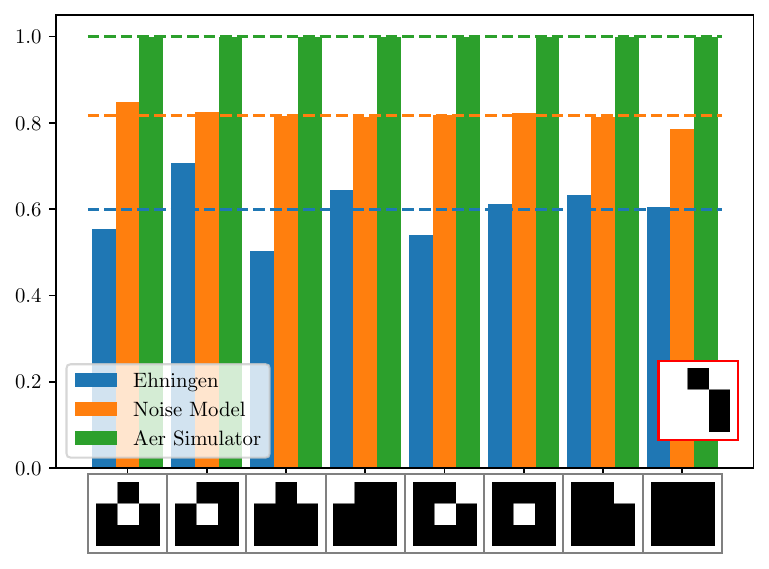}
         \caption{}
         \label{fig:result_plot_from_all_simulations:C}
     \end{subfigure}
     \begin{subfigure}[b]{0.49\linewidth}
         \centering
         \includegraphics[width=\textwidth]{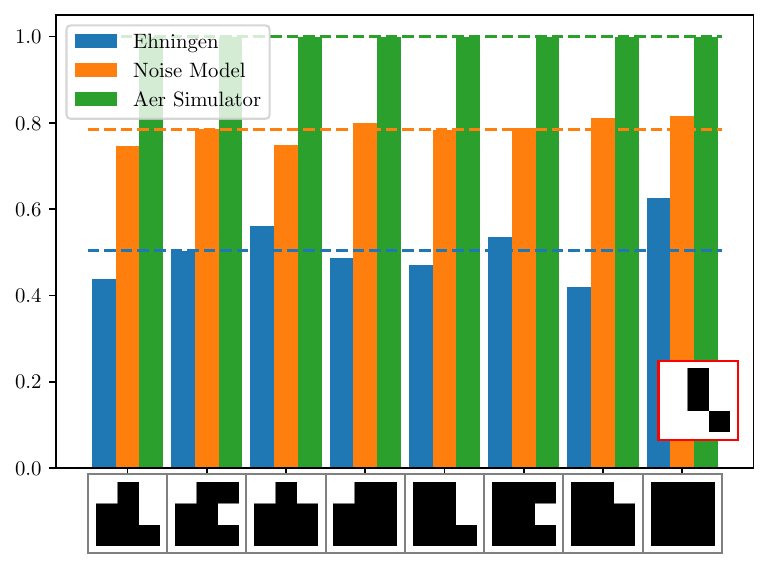}
         \caption{}
         \label{fig:result_plot_from_all_simulations:D}
     \end{subfigure}     
    \caption{The results from the run on the physical device in Ehningen is shown in blue. For comparison, the results from the simulation of the presented circuit with a backend model extracted from the original quantum computer in Ehningen (Noise Model, orange) and with an ideal QASM simulator (Aer Simulator, green) are presented as well. The x-axis shows all patterns which contain the target \textbf{A, B, C} or \textbf{D} to detect. On the y-axis, we plot the certainty of detecting this pattern: (\subref{fig:result_plot_from_all_simulations:A}) Shows the results for detecting pattern \textbf{A}, (\subref{fig:result_plot_from_all_simulations:B}) test pattern \textbf{B},  (\subref{fig:result_plot_from_all_simulations:C}) for test pattern \textbf{C} and (\subref{fig:result_plot_from_all_simulations:D}) for test pattern \textbf{D}. In the bottom right corner of each plot, the referenced target is shown.}
    \label{fig:result_plot_from_all_simulations}
\end{figure*}

Proper functioning of the circuit is ensured by Boolean quantum design rules. Pixels are placed only at odd $-\pi$ positions. \texttt{RZ} rotations are allowed only as multiples of $4\pi$. Pixels in the min-max range from the diagonal to the vertical bar are instantiated, and pixels below the diagonal remain blank. For the $3\times3$ PQHT case, there are 6 input pixels and the number of test vectors amounts to 64 for all pixel permutations. 
With help of IBM's aer-simulator, the full coverage cycle of 64 test input vectors has been applied to ensure the PQHT correctness: the aer-simulator outputted the correct results for all 64 PQHT test vectors. 
In the next step, 
%the PQHT Qiskit code of the composer has been exported and converted to an automated notebook script which transpiles the PQHT for each test vector 
we transpiled the quantum circuits 
for the target Ehningen quantum computer, using tool provided by Qiskit. For transpilation of the circuits, an optimization level of three has been adopted. Jobs have been created and run at the Ehningen quantum computer, with the output results contained in the histogram distributions for each test vector with various designs. At last, the one has been selected, which produced the best results in absolute max channel frequency. The image representation and the correctness of the orchestrated rotations have been shown and tested with the full coverage set of test vectors in the Ehningen quantum computer. Additionally, we extracted a backend model  Ehningen Quantum System One, capturing the noise of the physical system. This is used for comparison. 
From each (virtual) backend - the ideal AER QASM simulator, the noise backend model and the physical - a full test run with all possible input vectors have been executed. This has been done for 2048 shots in each of the simulators for reason of simulation time. In the physical system, 19999 shots have been used for maximal significance of the results. However, the statistic is expected to be valid despite the difference in the number of experiments. The results are presented together for comparison in \textbf{Fig.} \ref{fig:result_plot_from_all_simulations}.

\section{Job runs on the IBM Q System One in Ehningen}
Fraunhofer has an IBM Q System One installation in Ehningen, BW, Germany which can be remotely accessed and used. A four-month test time has been given to the authors to implement and test the PQHT during the year 2023. This time has been extensively used by numerous variations and job runs. The authors noticed the big difference between simulation and the unexpected results in real QC job runs. The simulation gives the correct results. For all 64 input vectors, the results coincide with the truth table.
%the truth tables and all four patterns are found correctly and the no operation switching conditions are correctly fulfilled as well.
For the six pixel involved, there are 64 pixel settings and their respective truth table is computed according to the existence of the set of four patterns to be identified. The criterion for a match is given, if the maximum histogram channel is the one which is expected from the truth table. The authors found the transpiler settings as the most critical issues for successful operation. The transpiler composes for each of the 64 pixel settings a transpiled quantum circuit. Many of the runs can't reproduce the truth table completely, because the transpiled circuits contain too many inserted SWAPs and the complexity is too high in terms of used quantum gates, even for that small $3\times3$ grid. 
This observation reveals the limited use of today's quantum computers. 
Still, we created a job (ID cj33trvrp5etsvercgs) %has been run and completed 
on Jul 30, 2023 12:43 at $ibmq\_ehningen$ with the transpiler instruction \texttt{tqc\_list = transpile(qc\_list, my\_backend, optimization\_level=3, seed\_transpiler=20)}. It reproduces the correct truth table for all 64 pixel settings to hundred per cent.

\section{Discussion}

The truth is in the programming details of the quantum computer. Simulation is necessary for quantum design verification, but only real runs on quantum computers warrant successful designs. Each design has to be tested and verified on the target quantum computer with the full test vector coverage. At this stage, the quantum computer sometimes remains an oracle, as various transpiled circuits have to be examined for their usability. The simulators give a general hint for the working of the circuit. With a simulator based on a more or less realistic noise model, the results come closer to the ones from the physical system. Even if the designed circuit works very well inside the simulation environment, it becomes difficult to obtain results with a similar certainty. Benchmarks should therefore be done in regard to the chosen quantum computer.

%In general, the results show a confidence better than 0.5, which is above the level of selecting a solution randomly out of all 8 possibilities. 

\section{Outlook}
The parallel quantum Hough transform consists of a matrix-like structure of parallel rows of sequentially connected \texttt{RZ} rotation operators and perpendicularly intertwined coincidence detector circuits. The rotations in \texttt{RZ} are precomputable to fit a predefined Hough feature space. The PQHT is inherently scalable in the number of input pixel Qubit lines \textit{n} and the number of search patterns \textit{m}. The $3\times3$ case expands to $n+1$ by adding more input Qubit lines and inserting additional \texttt{RZ} rotations in the Qubit lines. Accordingly, the coincidence unit is adapted to scalability.
However, the simple maxfinder unit needs to be replaced by scalable binary adders of size \textit{i}. Optionally, a threshold comparator
%of size \textit{l}
can be employed. 
As a field study, we implemented a 3 bit quantum binary adder and a 2 bit threshold comparator circuit. The 3 bit binary adder has been implemented by replicating the IBM composer design given by author P. Das \cite{uk20193bitbinary}. The 2 bit threshold comparator has been reconstructed in the IBM Composer from the circuit given by author H. Maity \cite{maity2022design}, where we only need the  PLA minterm $A < B$ for our threshold function. Global threshold values can be set by two comparator Qubit lines and connecting them to the comparison units. 
The 3 bit adder and the PLA minterm $A < B$  circuit have been verified separately with the full coverage set of all possible test vectors on the Ehningen quantum computer. 
The adder/comparator module can be scaled by extending the 3 bit binary adder to four or higher and building recursively adder trees thereof and extending the 2 bit comparator to 3 inputs or higher.
The speed-up achieved with the quantum circuit can be used to multiplex larger 2D data fields. This makes proper systems in the pre- and post-processing necessary, which is out of scope for this contribution. During our work on this algorithm, it has become clear, that an important bottleneck is the encoding of information for quantum systems. The classical logical encoding makes it difficult to obtain all the computing power available by quantum computers.

\bibliographystyle{quantum}
% \begin{thebibliography}{9}
% \bibitem{examplecitation}
%  Name Surname,
%  \href{https://doi.org/10.22331/
%        idonotexist}{Quantum
%        \textbf{123}, 123456 (1916).}

% \bibitem{biblatexsubmittingtothearxiv}
%  StackExchange discussion on \href{http://tex.stackexchange.com/questions/26990/biblatex-submitting-to-the-arxiv}{``Biblatex: submitting to the arXiv'' % (2017-01-10)}

%\bibitem{arxivpdfoutput}
%  Help article published by the arXiv on \href{https://arxiv.org/help/submit_tex}{``Considerations for TeX Submissions'' (2017-01-10)}

% \bibitem{howtogetdoilinksinbibliography}
%  StackExchange discussion on \href{http://tex.stackexchange.com/questions/3802/how-to-get-doi-links-in-bibliography}{``How to get DOI links in 
% bibliography'' (2016-11-18)}

% \bibitem{automaticallyaddingdoifieldstoahandmadebibliography}
%  StackExchange discussion on \href{http://tex.stackexchange.com/questions/6810/automatically-adding-doi-fields-to-a-hand-made-bibliography}
% {``Automatically adding DOI fields to a hand-made bibliography'' (2016-11-18)}
% \end{thebibliography}
\bibliography{references}

\begin{thebibliography}{10}

\bibitem{ballard1981generalizing}
Dana~H Ballard.
\newblock ``Generalizing the {H}ough transform to detect arbitrary shapes''.
\newblock \href{https://dx.doi.org/10.1016/0031-3203(81)90009-1}{Pattern recognition {\bf 13}, 111--122}~(1981).

\bibitem{mukhopadhyay2015survey}
Priyanka Mukhopadhyay and Bidyut~B Chaudhuri.
\newblock ``A survey of {H}ough {T}ransform''.
\newblock \href{https://dx.doi.org/10.1016/j.patcog.2014.08.027}{Pattern Recognition {\bf 48}, 993--1010}~(2015).

\bibitem{gautam2023image}
Sarita Gautam and Anuj Kumar.
\newblock ``Image-based automatic traffic lights detection system for autonomous cars: a review''.
\newblock \href{https://dx.doi.org/10.1007/s11042-023-14340-1}{Multimedia Tools and ApplicationsPages 1--48}~(2023).

\bibitem{rahman2020analysis}
S~Rahman, M~Ramli, F~Arnia, R~Muharar, M~Luthfi, and S~Sundari.
\newblock ``Analysis and comparison of {H}ough transform algorithms and feature detection to find available parking spaces''.
\newblock In Journal of Physics: Conference Series.
\newblock \href{https://dx.doi.org/10.1088/1742-6596/1566/1/012092}{Volume 1566, page 012092}.
\newblock IOP Publishing~(2020).

\bibitem{zakaria2023lane}
NJ~Zakaria, MI~Shapiai, RA~Ghani, MNM Yasin, MZ~Ibrahim, and N~Wahid.
\newblock ``Lane detection in autonomous vehicles: A systematic review''.
\newblock \href{https://dx.doi.org/10.1109/access.2023.3234442}{IEEE Access}~(2023).

\bibitem{seifozzakerini2018hough}
Sajjad Seifozzakerini, Wei-Yun Yau, Kezhi Mao, and Hossein Nejati.
\newblock ``{H}ough transform implementation for event-based systems: Concepts and challenges''.
\newblock \href{https://dx.doi.org/10.3389/fncom.2018.00103}{Frontiers in computational neuroscience {\bf 12}, 103}~(2018).

\bibitem{ayyad2023neuromorphic}
Abdulla Ayyad, Mohamad Halwani, Dewald Swart, Rajkumar Muthusamy, Fahad Almaskari, and Yahya Zweiri.
\newblock ``Neuromorphic vision based control for the precise positioning of robotic drilling systems''.
\newblock \href{https://dx.doi.org/10.1016/j.rcim.2022.102419}{Robotics and Computer-Integrated Manufacturing {\bf 79}, 102419}~(2023).

\bibitem{bachiller2018spiking}
Pilar Bachiller-Burgos, Luis~J Manso, and Pablo Bustos.
\newblock ``A spiking neural model of {HT3D} for corner detection''.
\newblock \href{https://dx.doi.org/10.3389/fncom.2018.00037}{Frontiers in Computational Neuroscience {\bf 12}, 37}~(2018).

\bibitem{vitale2021event}
Antonio Vitale, Alpha Renner, Celine Nauer, Davide Scaramuzza, and Yulia Sandamirskaya.
\newblock ``Event-driven vision and control for {UAV}s on a neuromorphic chip''.
\newblock In 2021 IEEE International Conference on Robotics and Automation (ICRA).
\newblock \href{https://dx.doi.org/10.1109/ICRA48506.2021.9560881}{Pages 103--109}.
\newblock IEEE~(2021).

\bibitem{kawakami2020cell}
Susumu Kawakami, Takehiro Ito, Yoshinari Makino, Makoto Hashimoto, and Masafumi Yano.
\newblock ``A cell model in the ventral visual pathway for the detection of circles of curvature constituting figures''.
\newblock \href{https://dx.doi.org/10.1016/j.heliyon.2020.e05397}{Heliyon {\bf 6}, e05397}~(2020).

\bibitem{laghari2012processor}
Mohammad~S Laghari and Gulzar~A Khuwaja.
\newblock ``Processor scheduling on parallel computers''.
\newblock In Proc. Int. Conf. on Computer, Electrical, and Systems Sciences, and Engineering, Abu Dhabi.
\newblock Citeseer~(2012).

\bibitem{park2019lane}
Hyunhee Park.
\newblock ``Lane detection algorithm based on {H}ough transform for high-speed self driving vehicles''.
\newblock \href{https://dx.doi.org/10.1504/IJWGS.2019.100835}{International Journal of Web and Grid Services {\bf 15}, 240--250}~(2019).

\bibitem{bailey2017streamed}
Donald~G Bailey.
\newblock ``Streamed {H}ough transform and line reconstruction on {FPGA}''.
\newblock In 2017 International Conference on Image and Vision Computing New Zealand (IVCNZ).
\newblock \href{https://dx.doi.org/10.1109/IVCNZ.2017.8402473}{Pages 1--6}.
\newblock IEEE~(2017).

\bibitem{256642}
F.~Klefenz, K.-H. Noffz, W.~Conen, R.~Zoz, A.~Kugel, and R.~Manner.
\newblock ``Track recognition in 4 $\mu$s by a systolic trigger processor using a parallel {H}ough transform''.
\newblock \href{https://dx.doi.org/10.1109/23.256642}{IEEE Transactions on Nuclear Science {\bf 40}, 688--691}~(1993).

\bibitem{DBLP:phd/dnb/Klefenz92}
Frank Klefenz.
\newblock ``{Ein systolischer Arrayprozessor zur Spurrekonstruktion als Trigger in Driftkammern}''.
\newblock PhD thesis.
\newblock University of Heidelberg, Germany.
\newblock ~(1992).
\newblock  url:~\url{https://d-nb.info/921412789}.

\bibitem{Klefenz91asystolic}
F.~Klefenz and R.~Männer.
\newblock ``{A Systolic Track Finding Trigger Processor}''.
\newblock In Proc. Comp. in High Energy Physics '91.
\newblock Pages 219--222.
\newblock ~(1991).

\bibitem{1003733}
A.~Epstein, G.U. Paul, B.~Vettermann, C.~Boulin, and F.~Klefenz.
\newblock ``A parallel systolic array {ASIC} for real-time execution of the {H}ough transform''.
\newblock \href{https://dx.doi.org/10.1109/TNS.2002.1003733}{IEEE Transactions on Nuclear Science {\bf 49}, 339--346}~(2002).

\bibitem{wang2021review}
Zhaobin Wang, Minzhe Xu, and Yaonan Zhang.
\newblock ``Review of quantum image processing''.
\newblock \href{https://dx.doi.org/10.1007/s11831-021-09599-2}{Archives of Computational Methods in EngineeringPages 1--25}~(2021).

\bibitem{yan2022toward}
Fei Yan, Salvador~E Venegas-Andraca, and Kaoru Hirota.
\newblock ``Toward implementing efficient image processing algorithms on quantum computers''.
\newblock \href{https://dx.doi.org/10.1007/s00500-021-06669-2}{Soft ComputingPages 1--13}~(2022).

\bibitem{amankwah2022quantum}
Mercy~G Amankwah, Daan Camps, E~Bethel, Roel Van~Beeumen, and Talita Perciano.
\newblock ``Quantum pixel representations and compression for {N-}dimensional images''.
\newblock \href{https://dx.doi.org/10.1038/s41598-022-11024-y}{Scientific reports {\bf 12}, 1--15}~(2022).

\bibitem{varadarajan2013quantum}
Karthik~Mahesh Varadarajan.
\newblock ``{Quantum Hough Transform}''.
\newblock In Graphicon'2013.
\newblock Pages 95--98.
\newblock ~(2013).

\bibitem{ma2021quantum}
Guangsheng Ma, Hongbo Li, and Jiman Zhao.
\newblock ``{Quantum Radon Transforms and Their Applications}''.
\newblock \href{https://dx.doi.org/10.1109/TQE.2021.3134648}{IEEE Transactions on Quantum Engineering {\bf 3}, 1--16}~(2021).

\bibitem{uk20193bitbinary}
Pratik Das.
\newblock ``Performing addition on {IBMs} quantum computers''.
\newblock https://quantumcomputinguk.org/tutorials\\/performing-addition-on-ibms-quantum-computers~(2019).

\bibitem{maity2022design}
Heranmoy Maity.
\newblock ``Design and implementation of a two-qubit quantum comparator circuit (q-cc)''.
\newblock \href{https://dx.doi.org/10.1007/s10825-022-01858-0}{Journal of Computational Electronics {\bf 21}, 530--534}~(2022).

\end{thebibliography}

\end{document}